\documentclass[aps,showpacs]{revtex4}
\usepackage{graphicx}

\begin{document}
\title{Numerical solutions to lattice-refined models in loop quantum cosmology}

\author{Subir Sabharwal and Gaurav Khanna}

\email{GKhanna@UMassD.Edu}

\affiliation{Physics Department, University of Massachusetts at Dartmouth, North Dartmouth, Massachusetts 02747}

\date{\today}
\begin{abstract}
In this article, we develop an intuitive and efficient, numerical technique to solve the quantum evolution equation of generic lattice-refined models in loop quantum cosmology. As an application of this method, we extensively study the solutions of the recently introduced, lattice-refined anisotropic model of the Schwarzschild interior geometry. Our calculations suggest that the results obtained from the approach are accurate, robust and are in complete agreement with the expectations from the von Neumann stability analysis of the model. 
\end{abstract}


\maketitle

\section{Introduction}
Loop quantum cosmology (LQC)~\cite{lqc} is a symmetry-reduced application of loop quantum gravity~\cite{rov}, a theory that leads to space-time that is discrete at Planck scales. An important result of LQC is that it is free of singularities. This is seen as a general consequence of the quantum evolution equation, which is a ``difference equation'' for the wave function and does not break down in the deep Planck regime, where the classical singularity is located~\cite{boj02}. 

Although, LQC models generically appear to exhibit good behavior on the Planck scale, some of these models run into problems in the semiclassical regime~\cite{Jess,Latest}. For example, in the semiclassical sector of the isotropic model with positive cosmological constant $\Lambda$, the Wheeler-DeWitt equation (which would be a close approximation to the difference equation of LQC) has solutions that oscillate on scales of $a\sqrt{\Lambda}$, where $a$ is the scale-factor. This scale becomes smaller with the expanding universe, ultimately becoming small enough that it falls under the discreteness scale of the LQC difference equation. Thus, such an LQC model would necessarily deviate from the correct semiclassical behavior~\cite{Latest}. Similar issues arise in the context of anisotropic LQC models, such as Bianchi I LRS~\cite{homo} and the Schwarzschild interior geometry~\cite{schwarz}, where the evidence of poor semiclassical behavior emerges in the form of a generic instability in the solutions of the quantum evolution difference equation~\cite{Jess}. 

These issues relating to the semiclassical behavior of certain LQC models can be resolved by introducing lattice-refinement~\cite{abhay}. The main concept of this modification is that the underlying lattice over which the quantum wave function has support, should be refined during the evolution of the model. In general this leads to a new type of the quantum evolution equation, one that no longer has a uniform step-size, thus presenting new challenges for its analysis and its solution. It should be emphasized that this modification is not necessary in the Planck regime, but only in semiclassical regime. This is because, near the classical singularity there are only a few action steps of the Hamiltonian that are relevant, and therefore lattice-refinement has little impact. Research on solving and analyzing lattice-refined difference equations is of current importance because it would play an important role in further advancing our understanding of semiclassical LQC and also help with the development of effective equations and dynamics. In fact, preliminary results from lattice-refined models of LQC suggest that it plays an important role in the development of an LQG based candidate for dark energy~\cite{dark} and also for the origin of cosmic inflation~\cite{nelson}. 

Based on full LQG, there is a heuristic way of understanding why lattice-refinement is essential. Consider the context of a cosmological model and consider a state of quantum geometry corresponding to a fiducial volume cell. As the scale-factor increases the number of vertices of the state (graph) would have to increase correspondingly. To compute the Hamiltonian constraint in the full theory, one would then need to calculate the holonomy around the faces of an elementary cube surrounding each vertex. Now, as the number of such elementary cubes contained in the specified cell grow with the universe, the area of their faces would decrease. And for this reason, the edge length which is used for the computation of the holonomy would decrease. Thus, the dependence of this edge length on the scale-factor can be thought of as an effect of the full theory on simpler cosmological models. This is the main idea behind lattice-refined models.      

In this work, we will present an intuitive and efficient numerical method for solving the Hamiltonian constraint of lattice-refined models in LQC. It should be noted that solving the Hamiltonian constraint with uniform discreteness structure i.e. without refinement, is relatively straightforward and there are a variety of analytic and numerical techniques available~\cite{mylqcstuff,Jess}. With lattice-refinement, the naive iterative approach towards computing a solution, simply fails. The reason is as follows: due to the non-uniform stepping, computing the future value of the wave function requires you to know past of values of the wave function, that are different from the ones computed in the previous iteration! We will explain this in detail with an example in the next section. Due to this difficulty it is simply not possible to use the large variety of techniques available in the numerical methods literature for solving such difference equations. However, in our current method we will overcome this issue through the intuitive approach of using a {\em local} interpolation formula for the solution, that will enable us to compute precisely those values of the wave function that are needed, to push the evolution forward. In other words, we will use the values of the wave function that were computed in the previous iteration to obtain a {\em local} analytic approximation formula of the solution. Then we will use this formula to compute the values of the wave function that will be needed for the next iteration. In this manner, we can obtain a complete iterative solution to the lattice-refined quantum evolution equation for any LQC model.  

This article is organized as follows. In the next section, through an explicit example, we will elaborate on the challenge associated with solving a $1D$ lattice-refined difference equation using a simple recursive scheme. We will then introduce our {\em local} interpolation based methodology and solve the same equation, effectively and efficiently. We will also solve the same relation using a different approach by performing a change of variable that will make the relation uniformly spaced (this can always be done in the context of $1D$ relations, but not in general) and then make an explicit comparison of the two solutions. In particular, we will demonstrate that they agree very well in the regime of interest. Next, we will proceed to solve a $2D$ lattice-refined LQC Hamiltonian constraint -- one that cannot be mapped over to an equi-spaced form -- using the same {\em local} interpolation based approach. We will demonstrate that the results in that case, agree with our general expectations of the solution, in particular, with regard to its stability properties. All the equations mentioned above, that will be solved in this work are relevant to the lattice-refined LQC model of the Schwarzschild interior geometry. We will end this article with a discussion of our results and some remarks on related work~\cite{kevin}.  

\section{Solving lattice-refined models}

We will start this section by very briefly reviewing the lattice-refined LQC model of the Schwarzschild interior geometry. We will follow the notation and treatment as presented in reference~\cite{Latest}. Consider a lattice with $N$ vertices that is adapted to the symmetry of the Schwarzschild interior geometry. If there are $N_\tau$ vertices along the direction of the triad labeled by $\tau$ and ${N_\mu}^2$ vertices in the spherical orbits of the symmetry group whose triad is labeled by $\mu$, then $N = {N_\tau} {N_\mu}^2$. Therefore, the step-size along $\mu$ would be $\delta/N_\mu$ and that along $\tau$ would be $\delta/N_\tau$. Following the process detailed in reference~\cite{Latest} we arrive at the following expression for the Hamiltonian constraint for this LQC model with lattice-refinement: 
\begin{eqnarray*}
&&2\delta N_{\mu}^{-1}(\sqrt{|\tau+2\delta N_{\tau}^{-1}|}+
\sqrt{|\tau|})
\left(\psi_{\mu+2\delta N_{\mu}^{-1},\tau+2\delta N_{\tau}^{-1}}- 
\psi_{\mu-2\delta N_{\mu}^{-1},\tau+2\delta N_{\tau}^{-1}}\right)
\nonumber\\
&& +(\sqrt{|\tau+\delta N_{\tau}^{-1}|}-\sqrt{|\tau-\delta N_{\tau}^{-1}|})
\left((\mu+2\delta N_{\mu}^{-1})\psi_{\mu+4\delta N_{\mu}^{-1},\tau}-
2(1+2\gamma^2\delta^2 N_{\mu}^{-2})\mu\psi_{\mu,\tau}\right.\nonumber\\
&&\qquad\qquad+\left(
(\mu-2\delta N_{\mu}^{-1})\psi_{\mu-4\delta N_{\mu}^{-1},\tau}\right)\nonumber\\
&&+2\delta N_{\mu}^{-1}(\sqrt{|\tau-2\delta N_{\tau}^{-1}|}+\sqrt{|\tau|})
\left(\psi_{\mu-2\delta N_{\mu}^{-1},\tau-2\delta N_{\tau}^{-1}}-
\psi_{\mu+2\delta N_{\mu}^{-1},\tau-2\delta N_{\tau}^{-1}}\right)\nonumber\\ 
&=& 0\, . 
\end{eqnarray*}
We now have to make further assumptions on how exactly the lattice spacing is changing with changing $\mu$ and $\tau$. The simplest case is that the number of vertices in each direction is proportional to the geometrical area of a transverse surface. This yields $N_{\tau}\propto \sqrt{|\tau|}$ and $N_{\mu}\propto \sqrt{|\mu|}$. We will start with a detailed analysis of this case in this section. An alternate choice for $N_\tau$ and $N_\mu$, which was also considered in~\cite{Latest} is based on the intuition that the number of vertices in a given direction is proportional to the geometric extension of that direction. The resulting difference equation is more complex, but has improved stability properties~\cite{Latest} and therefore it is much preferable as a model for further study. We will analyze and solve that version, later in this section. 

\subsection{One-dimensional lattice-refined relations}

In the simpler context of $N_{\tau}\propto \sqrt{|\tau|}$ and $N_{\mu}\propto \sqrt{|\mu|}$, the above master difference equation is variable separable, which makes it possible to rewrite it in the form of two $1D$ difference equations (one with $\mu$ as the independent variable and the other with $\tau$) that have non-uniform stepping. If we write the wave function $\psi_{\mu,\tau}=A_{\mu}B_{\tau}$ then these two ordinary difference equations take the form shown below:
\begin{equation}
(\sqrt{|\tau+2\delta N_{\tau}^{-1}|}+\sqrt{|\tau|})B_{\tau+2\delta N_{\tau}^{-1}} 
-(\sqrt{|\tau-2\delta N_{\tau}^{-1}|}+\sqrt{|\tau|})B_{\tau-2\delta N_{\tau}^{-1}} 
     = -\lambda (\sqrt{|\tau+\delta N_{\tau}^{-1}|}-\sqrt{|\tau-\delta N_{\tau}^{-1}|}) B_{\tau}
\label{separated}
\end{equation}
\begin{equation}
(\mu+2\delta N_{\mu}^{-1})A_{\mu+4\delta N_{\mu}^{-1}}-2(1+2\gamma^2\delta^2 N_{\mu}^{-2})\mu A_{\mu}
+(\mu-2\delta N_{\mu}^{-1})A_{\mu-4\delta N_{\mu}^{-1}}
= \lambda (A_{\mu+2\delta N_{\mu}^{-1}}-A_{\mu-2\delta N_{\mu}^{-1}}) 2\delta N_{\mu}^{-1}
\label{separated2}
\end{equation}
Here $\lambda$ is the separation constant. Now, consider the $B_\tau$ relation that appears above in equation (\ref{separated}). In order to solve it, we can try using a naive recursive method, but we will demonstrate here that such an approach will be unsuccessful. 

\begin{figure}[ht]
\center
\includegraphics[width=0.6\textwidth]{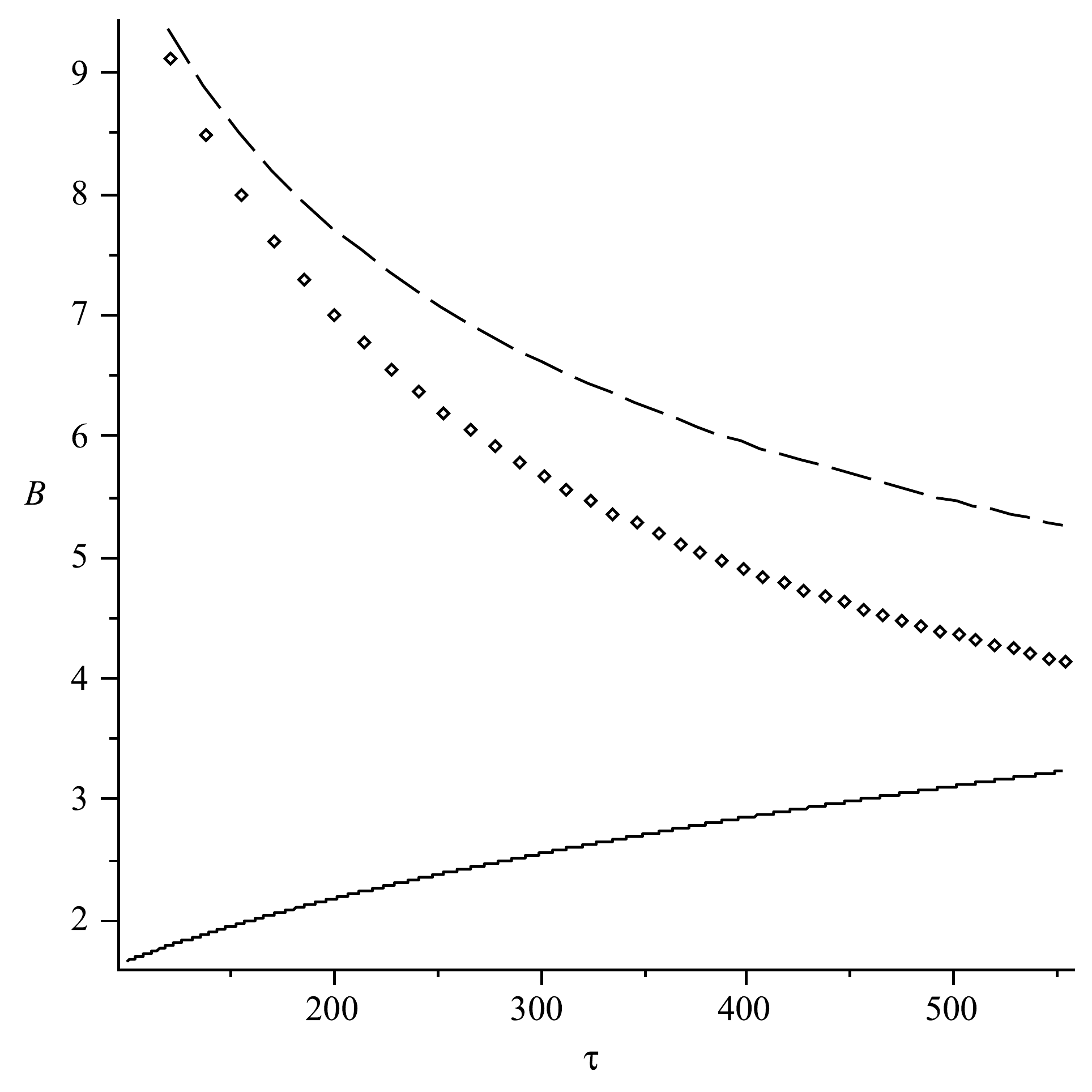}
\caption{\it We plot the solution of the $B_\tau$ relation for different values of the separation constant $\lambda$, using the  interpolation method as presented in the text. The dashed-line corresponds to a value of $\lambda=1.0$, the dotted line corresponds to $\lambda=2.1$ and the solid-line is for $\lambda=-5.1$. It is interesting to note that as the value of $\tau$ increases, all the plots start to appear like a solid and continuous line. This is because the density of the points on each plot increases with increasing $\tau$ (the stepping in $\tau$ decreases as $\delta/N_\tau$). }
\label{solution}
\end{figure}
\begin{figure}[ht]
\center
\includegraphics[width=0.6\textwidth]{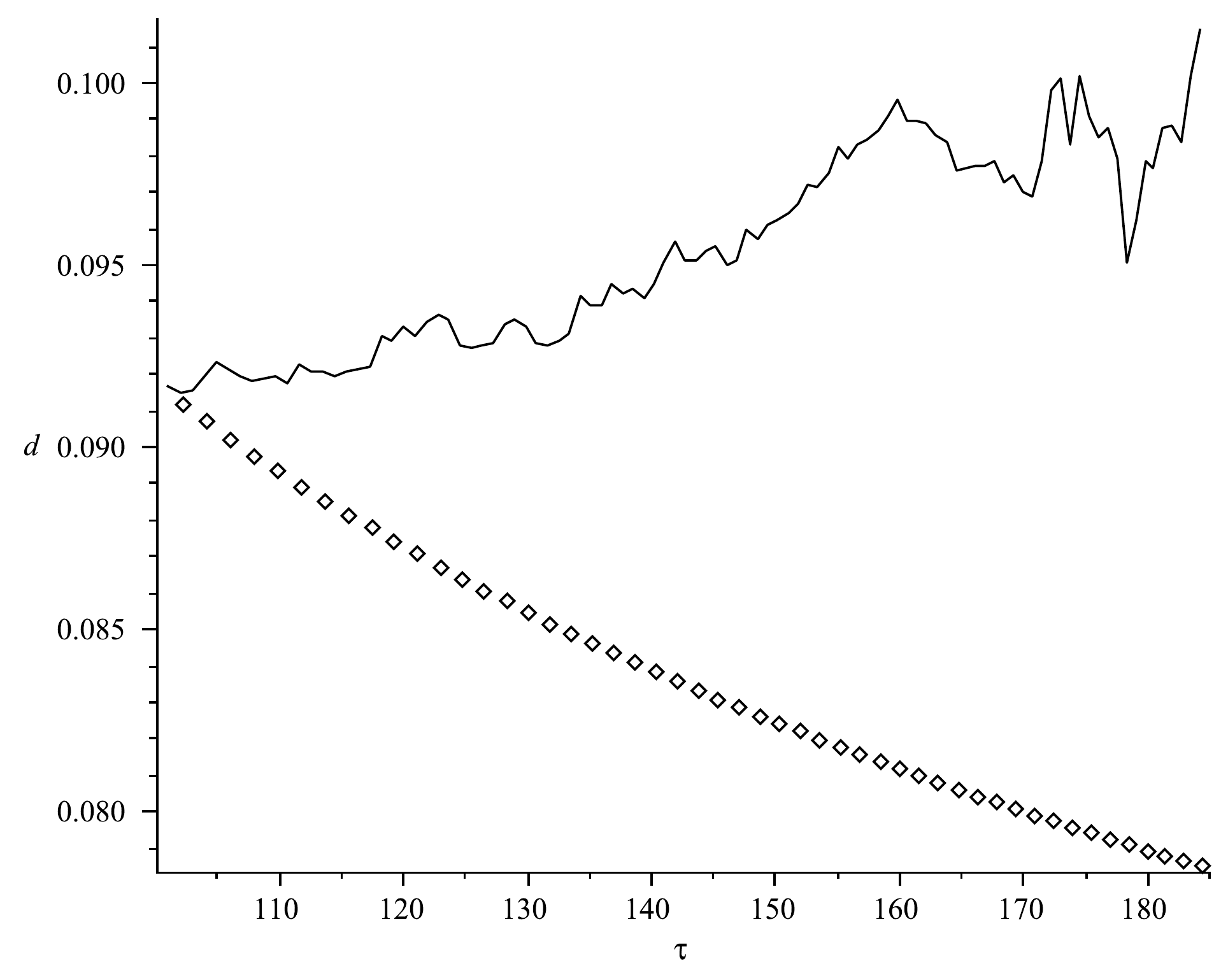}
\caption{\it We plot the percent-difference $d$, in the solutions of the $B_\tau$ relation for different types of interpolation schemes. The solid-line is the percent-difference between the solution based on the least-squares fit that we usually use, with the solution computed using a second-degree polynomial fit. The dotted-line is the percent-difference between the solution based on the least-squares fit, with the solution computed using a spline interpolation. The differences between these solutions are at the $0.1\%$ scale. This is a clear indication of our interpolation based method being very robust. }
\label{fits}
\end{figure}

To keep the discussion simple, assume $2\delta=1$. Lets start with some initial data for $B_\tau$: $B_{1-1/\sqrt{1}}=B_0=0$ and $B_1=1$. We have enough data to find the third point that is present in the $B_\tau$ relation (\ref{separated}) i.e., $B_{1+1/\sqrt{1}}=B_2$. Now, the parameter $\tau$ can advance only in steps of $1/\sqrt{\tau}$. Therefore, the next value that $\tau$ can take is $\tau+1/\sqrt{\tau}=1+1/\sqrt{1}=2$. With $\tau=2$, the $B_\tau$ relation has terms like $B_{2-1/\sqrt{2}}$, $B_2$ and $B_{2+1/\sqrt{2}}$. We just computed the value of $B_2$, but to use the relation above, in order to evaluate $B_{2+1/\sqrt{2}}$ we would now need to know the value of $B_{2-1/\sqrt{2}}$. However, there is no direct way of determining this value from the already known data i.e., $B_0$, $B_1$ and $B_2$. Thus, the recursive method of solving this difference equation seems to fail because the previous iteration is unable to provide us with data points needed for the next iteration. We resolve this problem in our method by performing a {\em local} interpolation to obtain an analytic formula using the values we already know, including the one we just computed from the relation itself. Then, we evaluate this interpolation formula at the points where the data values are needed in order to progress the evolution forward. More specifically, we proceed as follows: A least-squares fit is done with the points $B_0$, $B_1$ and $B_2$ to obtain a {\em locally} accurate formula for $B_\tau$. This is then used to evaluate $B_{2-1/\sqrt{2}}$ and thus, $B_{2+1/\sqrt{2}}$ can be evaluated. Now, this computed value is used to get a revised fit formula, that is accurate in the {\em local} neighborhood of $\tau =$ $2-1/\sqrt2$, $2$ and $2+1/\sqrt2$. Then, this revised formula is used to evaluate $B_\tau$ at the missing point i.e., $B_{{(2+1/\sqrt2)}-(1/\sqrt{2+1/\sqrt{2}})}$. Along with the value of $B_{2+1/\sqrt{2}}$, we can now compute $B_{{(2+1/\sqrt2)}+(1/\sqrt{2+1/\sqrt{2}})}$ and in this manner we can iterate and obtain the entire solution.

It should be noted that our interpolation based approach is geared towards finding only {\em pre-classical} solutions i.e. solutions that do not vary much between neighboring points. Restricting ourselves to pre-classical solutions in the context of this work is meaningful because our goal is to explore the semi-classical sector of LQC models. In fact, as pointed out earlier, lattice-refinement plays a significant role only in that regime.  

Complete solutions for the $B_\tau$ relation for a few different values of the separation constant $\lambda$ appear in Figure~\ref{solution}. An immediate concern that one may have with our approach is the matter of the sensitivity of our results to the choice of an  interpolation scheme i.e., the least-squares fit. In Figure~\ref{fits} we plot the percent-difference, $d$ between various solutions of the $B_\tau$ relation,  obtained by varying the type of interpolation scheme. The results clearly indicate that our solution method is very robust (the percent-difference is at the $0.1\%$ level). In the remaining portion of this article, we will continue to use a least-squares fit as our choice of the  interpolation scheme. 

\begin{figure}[ht]
\center
\includegraphics[width=0.6\textwidth]{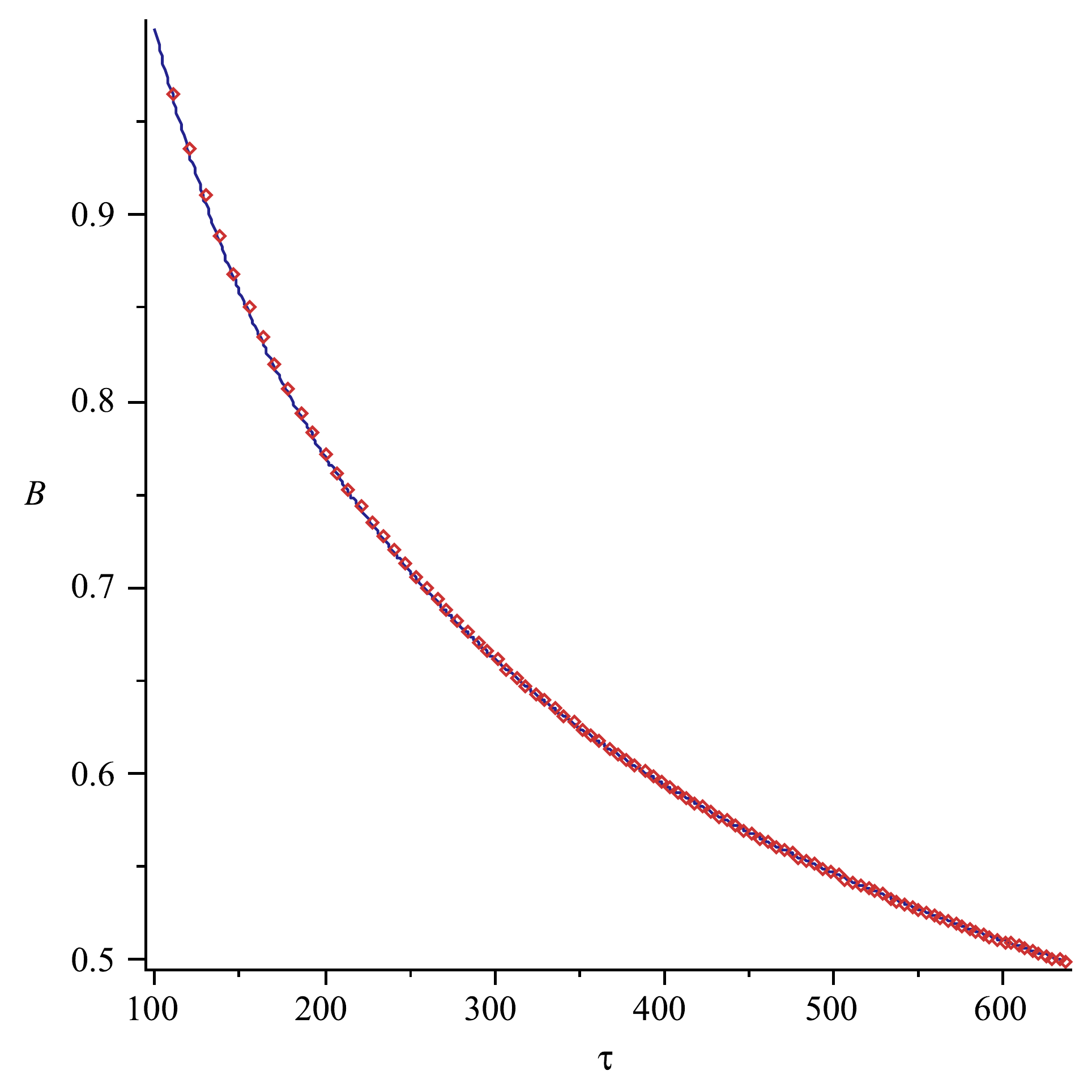}
\caption{\it We plot the solution of the $B_\tau$ relation solved using the interpolation approach alongside the approximate analytic solution of the same relation. Here the solid-line is the analytic solution. There is an excellent agreement between the two approaches.}
\label{comparison}
\end{figure}

An alternate approach that is applicable to all $1D$ lattice-refined relations~\cite{Latest,abhay}, suggests that we define a change of variable, $\bar{\tau} = {2\over 3} \tau^{3/2}$, one that can make the $B_\tau$ relation equi-spaced to linear order in $\delta$. This is because, in terms of this variable, $B_{\tau \pm 2\delta N_{\tau}^{-1}} \approx B_{\bar{\tau} \pm 2\delta}$ and then the $B_\tau$ relation (\ref{separated}), to linear order in $\delta$, becomes:
\begin{equation}
B_{\bar{\tau}+2\delta} - B_{\bar{\tau}-2\delta} \approx -(\lambda+2) B_{\bar\tau} \delta / 3 {\bar\tau}.
\end{equation}  
This particular relation has been studied before at various places in the literature~\cite{mylqcstuff} and its solutions are well known. It appears in the separable Bianchi I LRS (LQC, without lattice refinement) model and the asymptotic form of its solution can be written analytically as:
\begin{equation}
B_{\bar\tau} \propto {\bar\tau}^{-(\lambda+2)/12} \propto \tau^{-(\lambda+2)/8}.
\end{equation} 
In Figure~\ref{comparison} we plot the two solutions obtained using these different approaches together. The agreement is clearly seen to be excellent. 

\begin{figure}[ht]
\center
\includegraphics[width=0.6\textwidth]{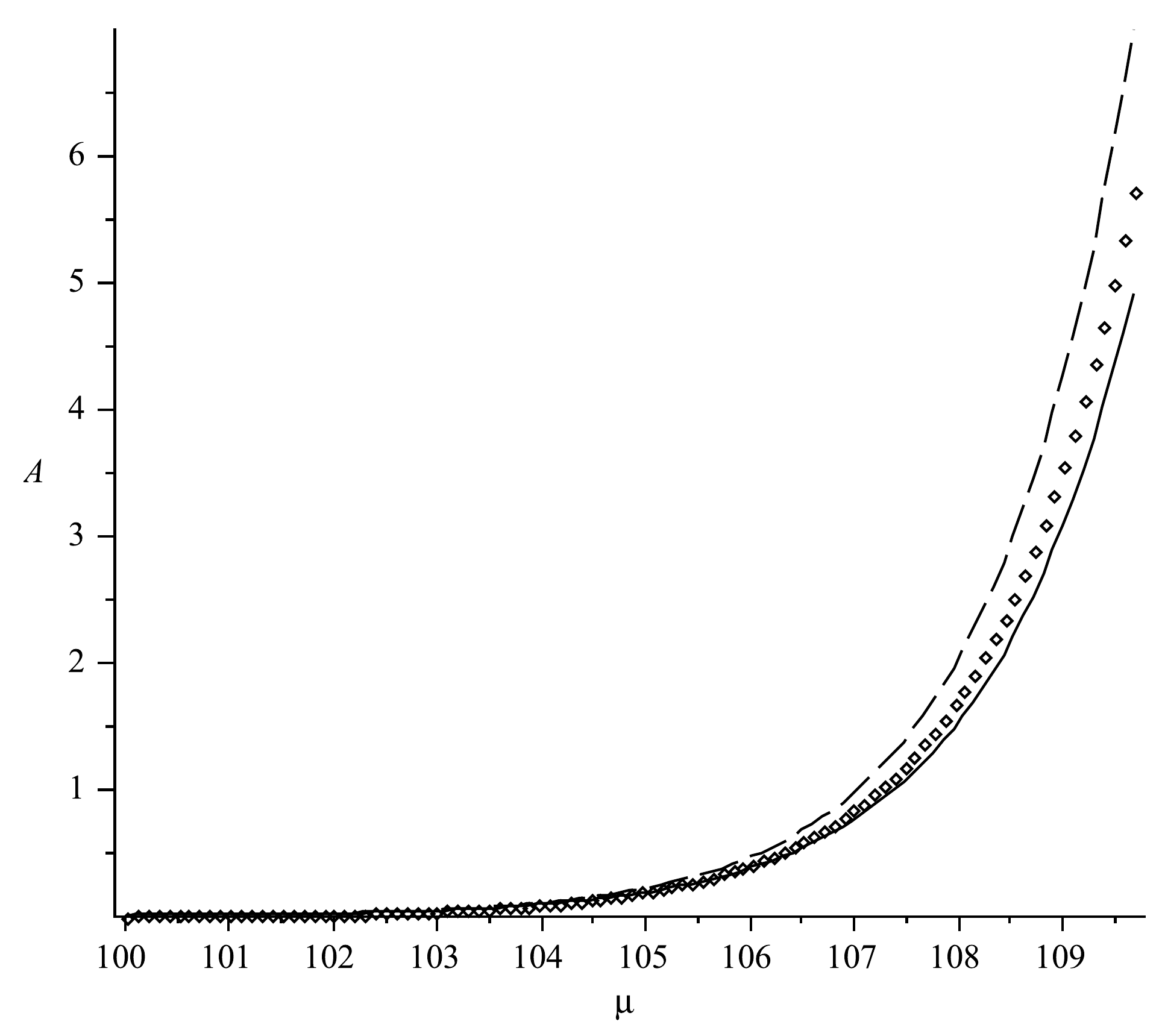}
\caption{\it We plot the solution of the $A_\mu$ relation for different values of the separation constant $\lambda$, using the  interpolation method as presented in the text. The dashed-line corresponds to a value of $\lambda=10$, the dotted line corresponds to $\lambda=5$ and the solid-line is for $\lambda=1$. }
\label{solution2}
\end{figure}

\begin{figure}[ht]
\center
\includegraphics[width=0.6\textwidth]{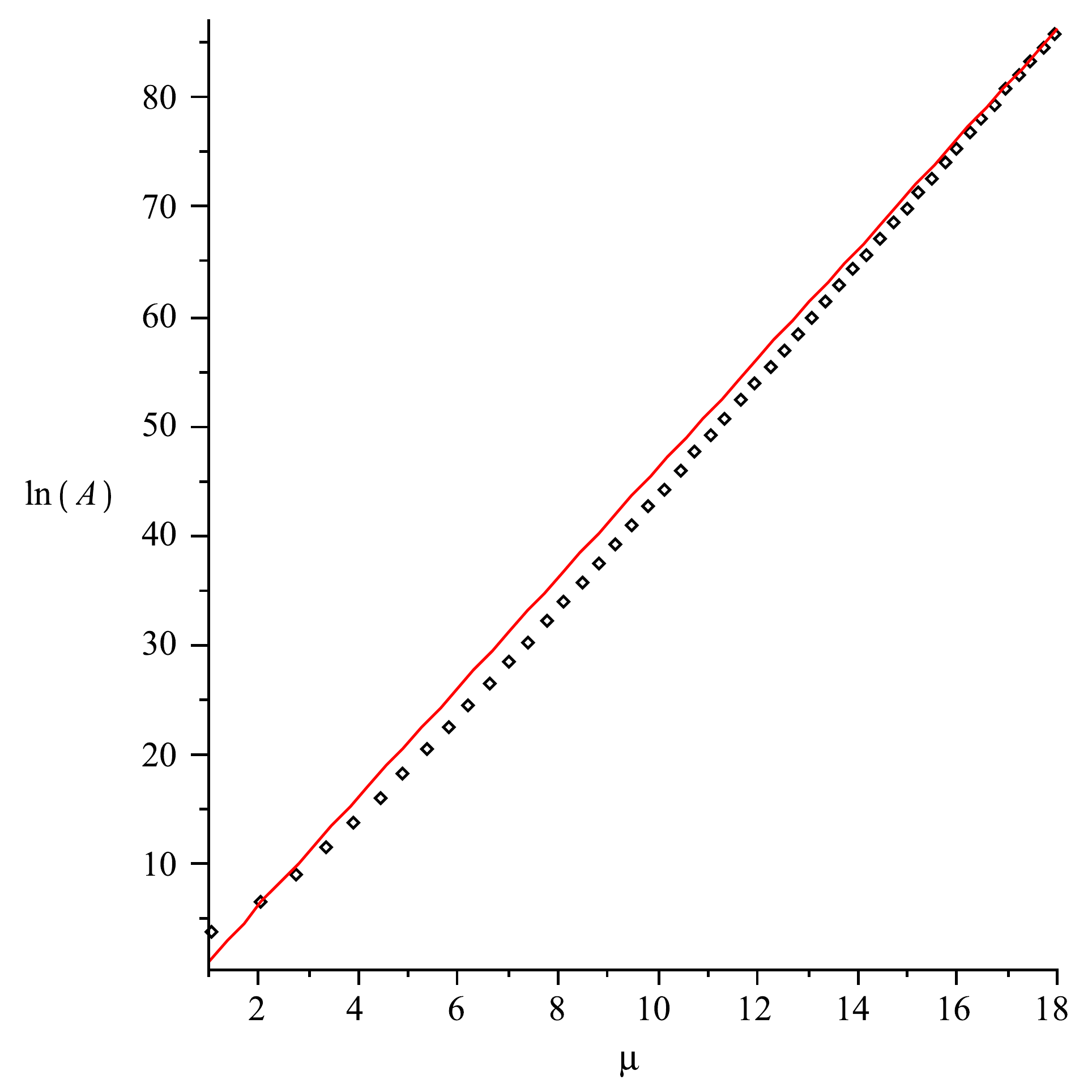}
\caption{\it We plot the logarithm of the solution of the $A_\mu$ relation solved using the interpolation approach alongside the approximate analytic solution of the same relation. Here the solid-line is the analytic solution. There is an excellent agreement between the two approaches, especially for large $\mu$.}
\label{comparison2}
\end{figure}

We now move on towards a solution of the $A_\mu$ relation (~\ref{separated2}). Note that our interpolation based method is immediately applicable to this relation as well. Indeed, this is the main merit of the technique; it is generically applicable to lattice-refined difference equations. Solutions for a few different values of the separation parameter $\lambda$ are plotted in Figure~\ref{solution2}. Let us proceed towards finding an asymptotic, analytic solution by transforming the relation to an equi-spaced one, using the new variable: $\bar{\mu} = {2\over 3} \mu^{3/2}$. Using, $A_{\mu \pm 4\delta N_{\mu}^{-1}} \approx A_{\bar{\mu} \pm 4\delta}$ and also that ${\bar\mu}>>\delta$ the $A_\mu$ relation can be simplified and written as an approximate ODE which takes the form shown below:
\begin{equation}
{\bar\mu} {{d^2}A\over{d{\bar\mu}}^{2}} - {\gamma^2}({{\bar\mu}\over 144})^{1/3} A=(\lambda-2)/3 {d A \over d{\bar\mu}}. 
\end{equation}
This simplified equation, that can be assumed to be valid asymptotically can be solved easily yielding  solutions in terms of modified Bessel functions: 
\begin{equation}
A_{\mu} \propto \mu^{(\lambda+1)/4} I_{(-\lambda-1)/4}(\gamma\mu/2),{\,}\mu^{(\lambda+1)/4} K_{(\lambda+1)/4}(\gamma\mu/2)   
\end{equation} 
It should be noted that these modified Bessel functions exhibit exponential behavior asymptotically, and it is related to this fact that the separable Schwarzschild model under consideration (with $N_{\tau}\propto \sqrt{|\tau|}$ and $N_{\mu}\propto \sqrt{|\mu|}$) exhibits generic instability~\cite{Latest}. In Figure~\ref{comparison2} we plot this analytic solution alongside the numerically generated solution. The agreement is clearly seen to be excellent.  


\subsection{Two-dimensional lattice-refined relations}

In the previous section, we solved the master Schwarzschild interior LQC relation for a simple intuitive choice for $N_\tau$ and $N_\mu$, one that makes the relation variable separable. We noted that the solutions to that version of the relation are unstable, which implies that the model cannot be viable as a physical model for the quantum geometry of the Schwarzschild black hole interior. This outcome is consistent with what was observed before~\cite{Latest} wherein, von Neumann numerical analysis of the same relation yielded the same unstable characteristics. Now, let us consider the alternate choice for $N_\tau$ and $N_\mu$, the one that has improved stability properties: $N_{\mu}\propto \sqrt{|\tau|}$ and $N_{\tau}\propto\mu/\sqrt{|\tau|}$. It is very interesting that the notion of von Neumann stability that plays ordinarily a significant role in numerical analysis of partial differential equations, provides an excellent  mechanism for limiting the wide range of possible choices for $N_\tau$ and $N_\mu$~\cite{Latest}.  

\begin{figure}[ht]
\center
\includegraphics[width=0.6\textwidth]{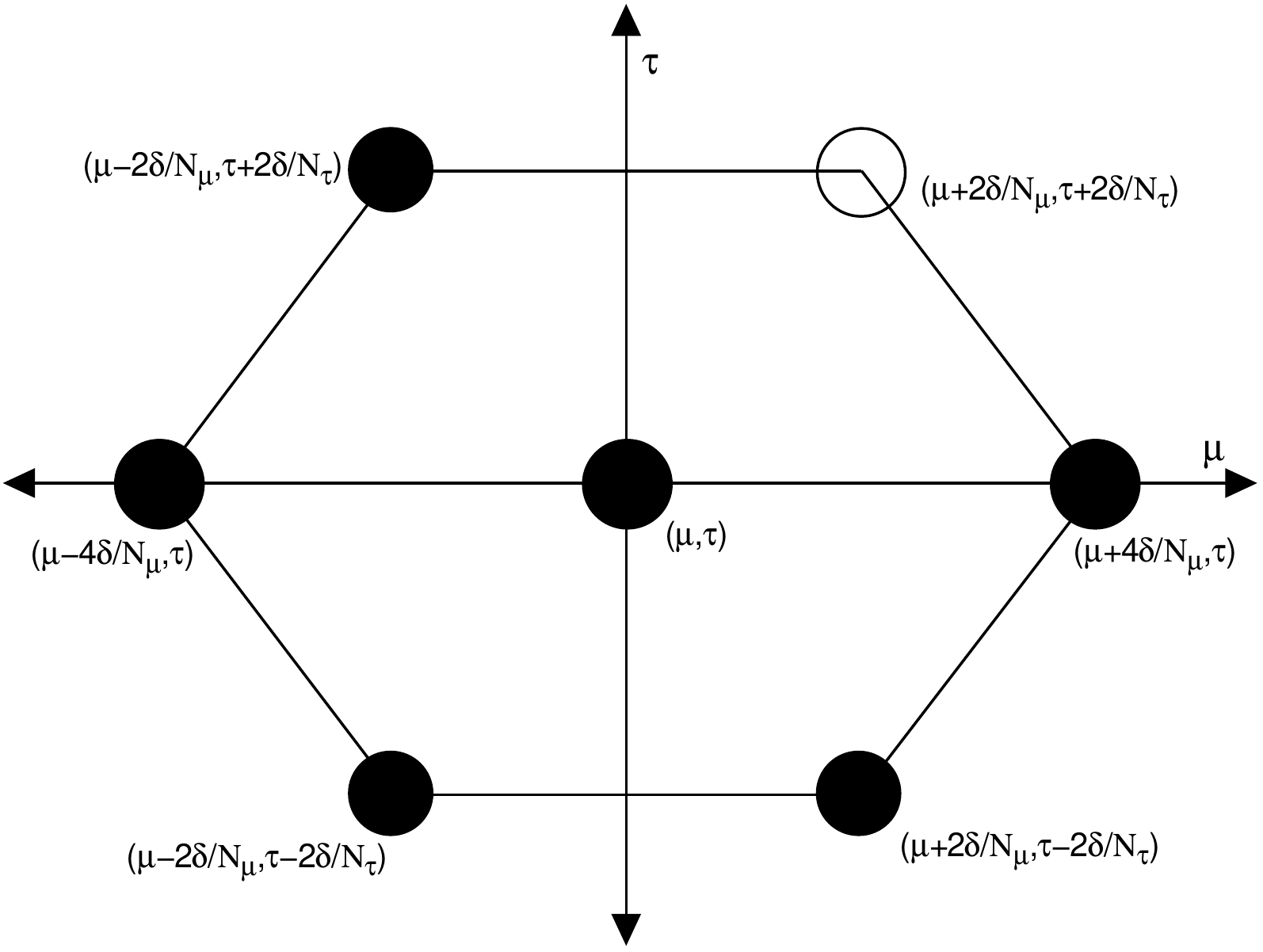}
\caption{\it The numerical ``stencil'' that is used for the evolution of a wave-packet in the lattice-refined LQC model of the Schwarzschild interior. This graphic depicts all the values of the wave function that are connected by the evolution difference equation (Hamiltonian constraint).   }
\label{stencil}
\end{figure}
\begin{figure}[ht]
\center
\includegraphics[width=0.6\textwidth]{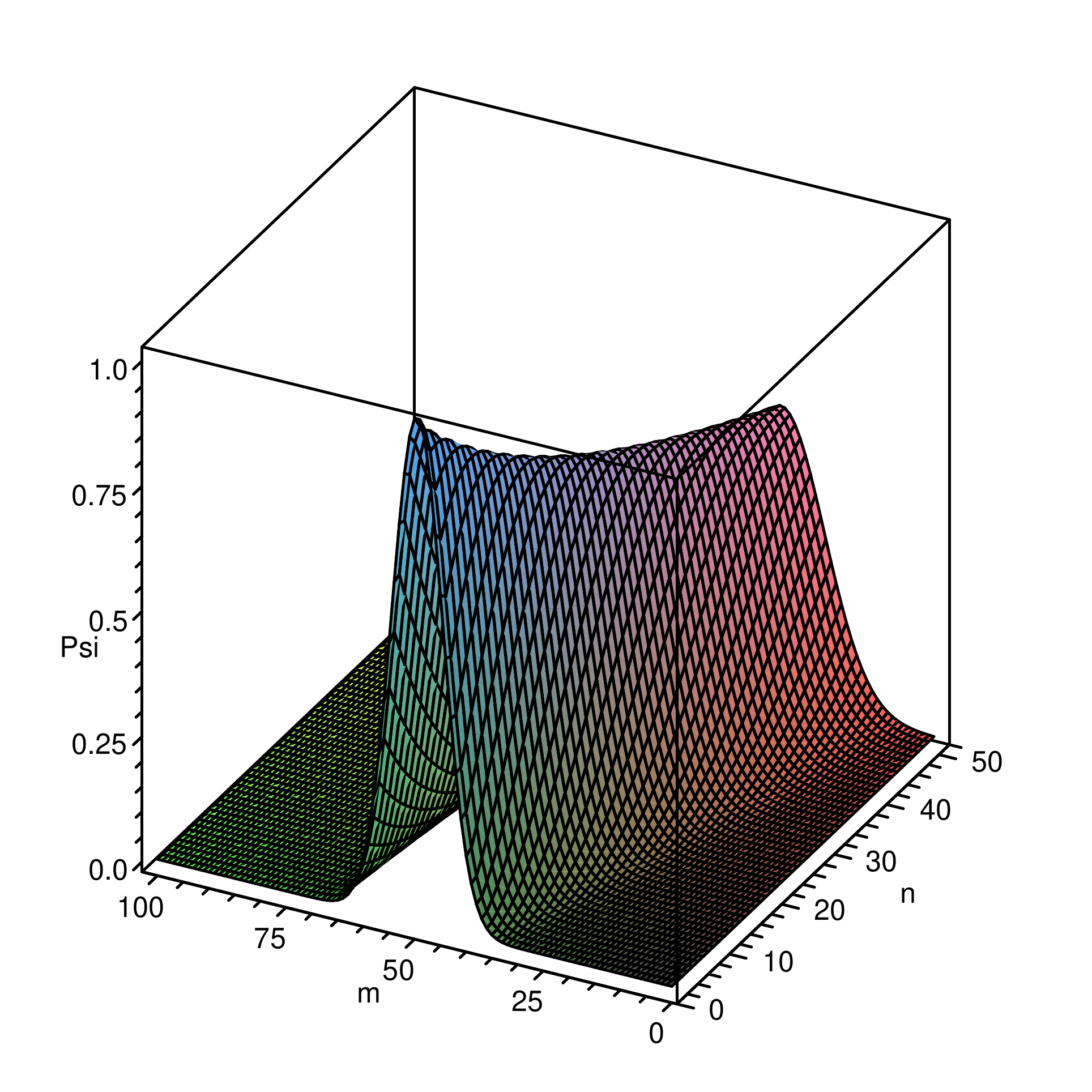}
\caption{\it Stable evolution of a gaussian wave-packet in the lattice-refined LQC model of the Schwarzschild interior, based on the local 2D interpolation method as presented in this article. The scales of the axis depicted above are based on $(m,n)$ and are arbitrary. The accompanying text has more detail. }
\label{stability}
\end{figure}

With this choice of $N_\tau$ and $N_\mu$, the relation under consideration is not variable separable. This is because the step-size of any of the two independent variables depends on the other variable. In addition, to complicate matters further, it is not possible to find two independent quantities that would make it possible for the difference equation to be transformed into a uniformly spaced version~\cite{Latest}, just like we did for the $1D$ relations in the previous section. Thus, we have no choice but to solve this non-uniformly spaced, partial difference equation directly. Fortunately, the {\em local} interpolation method, that we have introduced in this article and detailed in the previous section is readily applicable to this challenging problem.   

Conceptually, our approach towards solving this $2D$ lattice-refined difference equation is identical to the one we took in $1D$. The relation connects various points on the lattice, which are shown clearly in Figure~\ref{stencil}. In order to solve for the value $\psi_{\mu+2\delta N_{\mu}^{-1},\tau+2\delta N_{\tau}^{-1}}$ to evolve the solution forward, we need to have values for the wave function that in general, have not been computed from previous iterations. However, these values are in close proximity to values that are known from before, therefore it is very natural to perform a {\em local} $2D$ interpolation using the values that are known, to compute the values that are needed. To demonstrate the success of this approach, we evolve a semiclassical wave-packet and show the results in Figure~\ref{stability}. We explain the evolution process in some detail below.

We start by assuming a gaussian profile for the wave function $\psi$, and evolve it using the interpolation approach that we developed in the previous section. A $2D$ grid for $\mu$ and $\tau$ points is first set up by arbitrarily choosing some $(\mu,\tau)$ ordered pair with both $\mu$ and $\tau$ being large positive integers. The grid implements the fact that $\mu$ advances as $\mu+1/\sqrt{\tau}$ while $\tau$ advances as $\tau+\mu/\sqrt{\tau}$. Each point on the grid is labeled by two integers $(m,n)$ which would represent the corresponding position of the point on an uniformly spaced grid. Note that although we don't have an uniformly spaced grid, these labels help with ``book-keeping'' i.e. identifying the points being used for the iterations. The first row of points that have a constant $n$ label and are of the form $(\mu+1/\sqrt{\tau},\tau)$ i.e. the set of points (relevant to the relation) for which $\tau$ is a constant. All the other points evaluated on the grid are of the form $(\mu+1/\sqrt{\tau},\tau+\mu/\sqrt{\tau})$. Each of these points act as the central anchor for the other points that enter the relation. For example, referring to Figure~\ref{stencil}, it is the central point of the polygon that we evaluate while laying out the grid. All the points ``connected'' to it via the relation, can then be easily evaluated. 

\begin{figure}[ht]
\center
\includegraphics[width=0.6\textwidth]{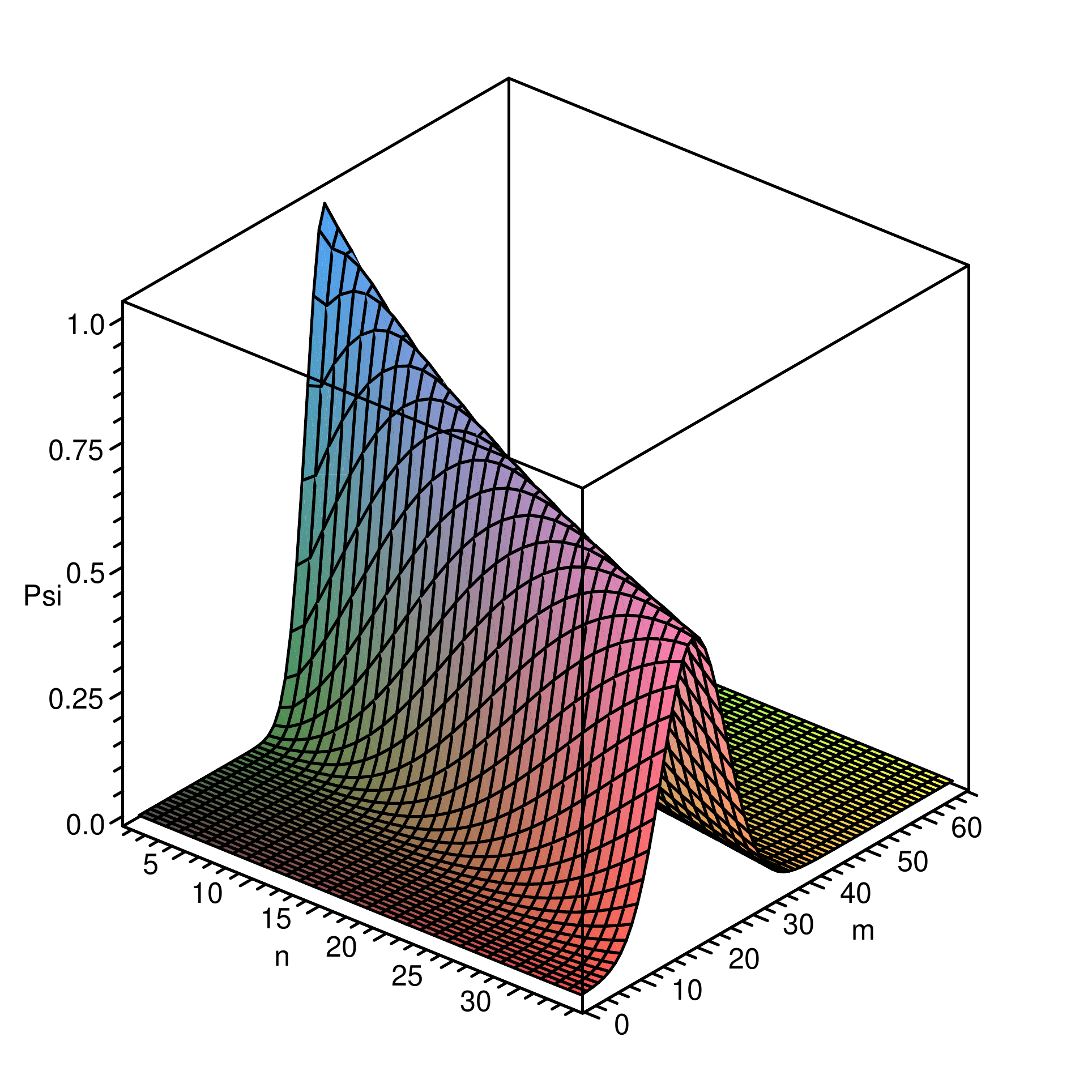}
\caption{\it Stable evolution of a gaussian wave-packet in the lattice-refined LQC model of Bianchi I LRS, based on the local 2D interpolation method as presented in this article. The scales on the axis depicted above are based on $(m,n)$ and are arbitrary. The accompanying text has more detail.}
\label{bianchi}
\end{figure}

Now that we have explained how the $2D$ grid is set up, let us turn the discussion towards evaluating and evolving the wave function at these discrete points. As mentioned before, the initial wave function is assumed to be a gaussian wave-packet centered on the $\mu$ axis. The values of the wave function at all the points marked as dark-circles in  Figure~\ref{stencil} are evaluated using the gaussian wave-packet expression and this data is used to find the value of the wave function at the location $(\mu+1/\sqrt{\tau},\tau+\mu/\sqrt{\tau})$, which is the white-circle in the figure. In this manner, we obtain the value of the function at points in the vicinity of $(\mu+1/\sqrt{\tau},\tau)$ for all $\mu$. In other words (with respect to labels $m,n$), we compute the values of the wave function for all values of $m$, for a given $n$. Then we move on to the next $n$-step (recall that $m$ and $n$ are just integer labels for the points with respect to the starting point, so they advance by one along their respective axes). But, now we find ourselves in the same situation as we were in $1D$ i.e. we do not have the data points we need to make the evolution progress further using the $2D$ master relation. Thus, we utilize a $2D$ least-squares fit to compute these unknown values using the data points known from the previous $n$-step, that have the same value of $m$. In other words, to find the value of the wave function that is needed by the $2D$ relation, we find a {\em local} $2D$ interpolation formula for points in the neighborhood of $(m,n)$ using the data from the $(m,n-1)$ point and the points around it (see Figure~\ref{stencil}). This yields an explicit scheme for progressing the evolution forward. 

The results from this approach are depicted in Figures~\ref{stability} (for the Schwarzschild interior model) and Figure~\ref{bianchi} (for the Bianchi I LRS model). It is clear that the evolution of the wave-packet is very smooth and stable as one would expect in the semi-classical regime. It should be noted for both figures, we chose the domain to be such that $\mu > 4\tau$, which happens to be the {\em unstable} region of the domain for the equi-spaced case~\cite{Jess}. Therefore, it is clearly evident that lattice-refinement has ``cured'' the problem of von Neumann instability in such anisotropic LQC models. This is completely consistent with the analytical stability analysis that was performed in~\cite{Latest}. The starting values we used for $(\mu,\tau)$ were $(100000,25000)$ and we performed approximately $200$ iterations in both directions ($250 \times 150$ grid points). In the near future we plan on developing a high-performance numerical code based on this current {\em proof-of-concept} approach that will be able to compute on much larger grid sizes, in a very effective manner. Then we will be able to study the semi-classical wave-packet trajectories on a much larger domain and compare them with the recently published effective equations and effective dynamics~\cite{kevin}. 

\section{Summary \& Discussion}

In this article, we have presented an effective numerical method for solving the Hamiltonian constraint in generic LQC models with lattice-refinement. The method is based on performing a {\em local} interpolation to obtain needed values from known values. 

We demonstrated the success of the method by solving the lattice-refined LQC model of the Schwarzschild interior geometry using two different refinement models. In the first refinement model, the $2D$ quantum evolution equation becomes variable-separable, reducing down to two $1D$ lattice-refined equations that we then solved using our interpolation technique and also analytically, using asymptotic approximations. Using this model as a test case, we compared the accuracy and the robustness of our interpolation approach. Then we chose a refinement model that is inherently non-separable, and solved that in the context of the evolution of semi-classical wave-packets using an extension of the same interpolation approach in two dimensions. The results from our evolutions reinforced recently published results on the von Neumann stability analysis of the lattice-refined Schwarzschild relation~\cite{Latest}. 

One preliminary remark that we would like make is that, based on our current results we do not see any evidence of the kind of behavior depicted in~\cite{kevin} which was obtained from a study of the effective semi-classical dynamics. In particular, in~\cite{kevin} the effective semi-classical Hamiltonian constraint is solved for the Schwarzschild interior model and those results suggest that: (a) In the equi-spaced model, there is a ``bounce'' and the quantum dynamics matches the solution to a {\em different} black hole (note that the ``bounce'' is not {\em through} the classical singularity, it actually {\em avoids} it!); and (b) In the lattice-refined case, one obtains an equilibrium solution asymptotically, thus {\em avoiding} the singularity (again). Results from the explicit evolution of wave-packets using the quantum Hamiltonian constraint appear to be qualitatively different. For example, in the equi-spaced case, past results~\cite{mylqcstuff,Jess} suggest that the quantum evolution equation allows for an evolution right {\em through} the classical singularity, as opposed to {\em avoiding} it. In addition, our preliminary results from the lattice-refined case show no indication of an asymptotic equilibrium, again as suggested by~\cite{kevin}. However, we make these preliminary remarks with caution. There are many caveats associated to the results presented in~\cite{kevin}, as the authors themselves point out (basically, they include one quantum effect, but ignore others). In addition, the approach towards numerical evolutions of wave-packets in lattice-refined LQC that we are presenting in this article is just a proof-of-concept demonstration, and is currently lacking any serious analysis and study. We will attempt to explore this apparent ``discrepancy'' in the near future.  

\section{Acknowledgments}

We thank Martin Bojowald and Daniel Cartin for many useful discussions related to this work, and for their  comments on an early draft of this article. We both acknowledge research support from the University of Massachusetts - Dartmouth. In addition, GK is supported by a grant from Glaser Trust of New York and SS is supported by various grants from the UMD Foundation.

\end{document}